\documentstyle[12pt]{article}
\def\doit#1#2{\ifcase#1\or#2\fi}

\skewchar\fivmi='177
\skewchar\sixmi='177
\skewchar\sevmi='177
\skewchar\egtmi='177
\skewchar\ninmi='177
\skewchar\tenmi='177
\skewchar\elvmi='177
\skewchar\twlmi='177
\skewchar\frtnmi='177
\skewchar\svtnmi='177
\skewchar\twtymi='177
\def\@magscale#1{ scaled \magstep #1}

\def\framingfonts#1{
\doit{#1}{\font\twfvmi  = ammi10   \@magscale5 % math italic
\skewchar\twfvmi='177
\skewchar\fivsy='60
\skewchar\sixsy='60
\skewchar\sevsy='60
\skewchar\egtsy='60
\skewchar\ninsy='60
\skewchar\tensy='60
\skewchar\elvsy='60
\skewchar\twlsy='60
\skewchar\frtnsy='60
\skewchar\svtnsy='60
\skewchar\twtysy='60
\font\twfvsy  = amsy10   \@magscale5 % math symbols
\skewchar\twfvsy='60
\font\go=font018			% Gothic
\font\sc=font005			% script
\def\Go#1{{\hbox{\go #1}}}	% Gothic for single characters in equations
\def\Sc#1{{\hbox{\sc #1}}}	% script for single characters in equations
\def\Sf#1{{\hbox{\sf #1}}}	% sans serif for single characters in equations
\font\oo=circlew10	      % thick circles (hollow: ` and a-n , solid:  p-z)
\font\ooo=circle10			% thin circles
\font\ro=manfnt				% font with rope
\def\kcl{{\hbox{\ro 6}}}		% left-handed rope
\def\kcr{{\hbox{\ro 7}}}		% right-handed rope
\def\ktl{{\hbox{\ro \char'134}}}	% top end for left-handed rope
\def\ktr{{\hbox{\ro \char'135}}}	% " right
\def\kbl{{\hbox{\ro \char'136}}}	% " bottom left
\def\kbr{{\hbox{\ro \char'137}}}	% " right
}}

\catcode`@=11
\def\un#1{\relax\ifmmode\@@underline#1\else
	$\@@underline{\hbox{#1}}$\relax\fi}
\catcode`@=12

\let\du=\d			% dot-under
\def\a{\alpha}

\def\d{\delta}
\def\e{\epsilon}

\def\g{\gamma}

\def\k{\kappa}
\def\l{\lambda}
\def\m{\mu}
\def\n{\nu}
\def\o{\omega}

\def\r{\rho}
\def\s{\sigma}

\def\z{\zeta}

\def\bo{{\raise.15ex\hbox{\large$\Box$}}}		% D'Alembertian
\def\pr{\prod}						% product
\def\TH{{\raise.2ex\hbox{$\displaystyle \bigodot$}\mskip-4.7mu \llap H \;}}
\def\face{{\raise.2ex\hbox{$\displaystyle \bigodot$}\mskip-2.2mu \llap {$\ddot
	\smile$}}}					% happy face
\def\sp#1{{}^{#1}}				% superscript (unaligned)
\def\Tilde#1{{\widetilde{#1}}\hskip 0.03in}			% big tilde
\def\Hat#1{\widehat{#1}}			% big hat
\def\Bar#1{\overline{#1}}			% big bar
\def\leftrightarrowfill{$\mathsurround=0pt \mathord\leftarrow \mkern-6mu
	\cleaders\hbox{$\mkern-2mu \mathord- \mkern-2mu$}\hfill
	\mkern-6mu \mathord\rightarrow$}
\def\dvec#1{\vbox{\ialign{##\crcr
	\leftrightarrowfill\crcr\noalign{\kern-1pt\nointerlineskip}
	$\hfil\displaystyle{#1}\hfil$\crcr}}}		% <--> accent
\def\frac#1#2{{\textstyle{#1\over\vphantom2\smash{\raise.20ex
	\hbox{$\scriptstyle{#2}$}}}}}			% fraction
\def\sfrac#1#2{{\vphantom1\smash{\lower.5ex\hbox{\small$#1$}}\over
	\vphantom1\smash{\raise.4ex\hbox{\small$#2$}}}}	% alternate fraction
\def\bfrac#1#2{{\vphantom1\smash{\lower.5ex\hbox{$#1$}}\over
	\vphantom1\smash{\raise.3ex\hbox{$#2$}}}}	% "
\def\afrac#1#2{{\vphantom1\smash{\lower.5ex\hbox{$#1$}}\over#2}}    % "
\newskip\humongous \humongous=0pt plus 1000pt minus 1000pt
\def\caja{\mathsurround=0pt}
\def\eqalign#1{\,\vcenter{\openup2\jot \caja
	\ialign{\strut \hfil$\displaystyle{##}$&$
	\displaystyle{{}##}$\hfil\crcr#1\crcr}}\,}
\newif\ifdtup
\def\panorama{\global\dtuptrue \openup2\jot \caja
	\everycr{\noalign{\ifdtup \global\dtupfalse
	\vskip-\lineskiplimit \vskip\normallineskiplimit
	\else \penalty\interdisplaylinepenalty \fi}}}
\def\li#1{\panorama \tabskip=\humongous				% eqalignno
	\halign to\displaywidth{\hfil$\displaystyle{##}$
	\tabskip=0pt&$\displaystyle{{}##}$\hfil
	\tabskip=\humongous&\llap{$##$}\tabskip=0pt
	\crcr#1\crcr}}

\doit0{
\def\ref#1{$\sp{#1)}$}
}

\parskip=\medskipamount			% space between paragraphs (LaTeX)
\thicklines			    % thick straight lines for pictures (LaTeX)

\def\border{						% border
	\setlength{\unitlength}{1mm}
	\newcount\xco
	\newcount\yco
	\xco=-24
	\yco=12
	\begin{picture}(140,0)
	\put(\xco,\yco){$\ktl$}
	\advance\yco by-1
	{\loop
	\put(\xco,\yco){$\kcl$}
	\advance\yco by-2
	\ifnum\yco>-240
	\repeat
	\put(\xco,\yco){$\kbl$}}
	\xco=158
	\yco=12
	\put(\xco,\yco){$\ktr$}
	\advance\yco by-1
	{\loop
	\put(\xco,\yco){$\kcr$}
	\advance\yco by-2
	\ifnum\yco>-240
	\repeat
	\put(\xco,\yco){$\kbr$}}
        \put(-20,11){\tiny University of Maryland Elementary Particle
Physics University of Maryland Elementary Particle Physics University of
Maryland Elementary Particle Physics}
	\put(-20,-241.5){\tiny University of Maryland Elementary
Particle Physics University of Maryland Elementary Particle Physics
University of Maryland Elementary Particle Physics}
	\end{picture}
	\par\vskip-8mm}
\def\bordero{						% alternate border
	\setlength{\unitlength}{1mm}
	\newcount\xco
	\newcount\yco
	\xco=-24
	\yco=12
	\begin{picture}(140,0)
	\put(\xco,\yco){$\ktl$}
	\advance\yco by-1
	{\loop
	\put(\xco,\yco){$\kcl$}
	\advance\yco by-2
	\ifnum\yco>-240
	\repeat
	\put(\xco,\yco){$\kbl$}}
	\xco=158
	\yco=12
	\put(\xco,\yco){$\ktr$}
	\advance\yco by-1
	{\loop
	\put(\xco,\yco){$\kcr$}
	\advance\yco by-2
	\ifnum\yco>-240
	\repeat
	\put(\xco,\yco){$\kbr$}}
	\put(-20,12){\ooo
bacdefghidfghghdhededbihdgdfdfhhdheidhdhebaaahjhhdahbahgdedgehgfdiehhgdigicba}
	\put(-20,-241.5){\ooo
ababaighefdbfghgeahgdfgafagihdidihiidhiagfedhadbfdecdcdfagdcbhaddhbgfchbgfdacfediacbabab}
	\end{picture}
	\par\vskip-8mm}
\def\headpic{						% UM heading
	\indent
	\setlength{\unitlength}{.4mm}
	\thinlines
	\par
	\begin{picture}(29,16)
	\put(165,16){\line(1,0){4}}
	\put(170,16){\line(1,0){4}}
	\put(180,16){\line(1,0){4}}
	\put(175,0){\line(1,0){4}}
	\put(180,0){\line(1,0){4}}
	\put(185,0){\line(1,0){4}}
	\put(169,0){\line(0,1){16}}
	\put(170,0){\line(0,1){16}}
	\put(179,0){\line(0,1){16}}
	\put(180,0){\line(0,1){16}}
	\put(184,0){\line(0,1){16}}
	\put(185,0){\line(0,1){16}}
	\put(169,16){\oval(8,32)[bl]}
	\put(170,16){\oval(8,32)[br]}
	\put(179,0){\oval(8,32)[tl]}
	\put(185,0){\oval(8,32)[tr]}
	\end{picture}
	\par\vskip-6.5mm
	\thicklines}
\def\title#1#2#3#4{\border\headpic {\hbox to\hsize{#4 \hfill UMDEPP #3}}\par
	\begin{center} \vglue .5in {\large\bf #1}\\[.6in]
	{#2}\\[.1in] {\it Department of Physics and Astronomy}\\
	{\it University of Maryland, College Park, MD 20742}\\[1.5in]
	{\bf Abstract}\\[.1in] \end{center} \begin{quotation}}	% title stuff
\def\Title#1#2#3#4#5#6#7{\border\headpic
	{\hbox to\hsize{#7 \hfill UMDEPP #6}}\par
	\begin{center} \vglue .4in {\large\bf #1}\\[.4in]
	{#2}\\[.1in] {\it Department of Physics and Astronomy}\\
	{\it University of Maryland, College Park, MD 20742}\\[.1in]
	{#3}\\[.1in] {\it {#4}}\\ {\it {#5}}\\[.5in] {\bf Abstract}\\[.1in]
	\end{center} \begin{quotation}}			% " for 2 authors
\def\endtitle{\end{quotation}\newpage}			% end title page

\def\sect#1{\bigskip\medskip \goodbreak \noindent{\bf {#1}} \nobreak \medskip}
\def\refs{\sect{References} \footnotesize \frenchspacing \parskip=0pt}
\def\Item{\par\hang\textindent}

\def\doit#1#2{\ifcase#1\or#2\fi}
\def\[{\lfloor{\hskip 0.35pt}\!\!\!\lceil\,}
\def\]{\,\rfloor{\hskip 0.35pt}\!\!\!\rceil}

\def\Lag{{\cal L}}
\def\du#1#2{_{#1}{}^{#2}}

\def\plpl{{+\!\!\!\!\!{\hskip 0.009in}{\raise -1.0pt\hbox{$_+$}}
{\hskip 0.0008in}}}
\def\mimi{{-\!\!\!\!\!{\hskip 0.009in}{\raise -1.0pt\hbox{$_-$}}
{\hskip 0.0008in}}}

\def\pl#1#2#3{Phys.~Lett.~{\bf {#1}B} (19{#2}) #3}
\def\np#1#2#3{Nucl.~Phys.~{\bf B{#1}} (19{#2}) #3}

\def\pr#1#2#3{Phys.~Rev.~{\bf D{#1}} (19{#2}) #3}

\def\ap#1#2#3{Ann.~of Phys.~{\bf {#1}} (19{#2}) #3}
\def\prep#1#2#3{Phys.~Rep.~{\bf {#1}C} (19{#2}) #3}

\def\ijmp#1#2#3{Int.~Jour.~Mod.~Phys.~{\bf A{#1}} (19{#2}) #3}

\def\eqques{{~\,={\hskip -11.5pt}\raise -1.8pt\hbox{\large ?}
{\hskip 4.5pt}\,}}

\def\fracmm#1#2{{{#1}\over{#2}}}

\def\half{{\fracm12}}

\def\frac#1#2{{\textstyle{#1\over\vphantom2\smash{\raise -.20ex
	\hbox{$\scriptstyle{#2}$}}}}}			% fraction

\def\fracm#1#2{\hbox{\large{${\frac{{#1}}{{#2}}}$}}}

\def\Tilde#1{{\widetilde{#1}}\hskip 0.015in}
\def\Hat#1{\widehat{#1}}
\def\scst{\scriptstyle}

\def\.{.$\,$}

\def\un{\underline}
\def\-{{\hskip 1.5pt}\hbox{-}}

\def\footnotew#1{\footnote{\hsize=6.5in {#1}}}

\def\low#1{{\raise -3pt\hbox{${\hskip 1.0pt}\!_{#1}$}}}

\begin{document}
% ---------------- end of def.tex ---------------

% -------------- Additional Fonts -----------------------
\font\tenmib=cmmib10
\font\sevenmib=cmmib10 at 7pt % =cmmib7 % if you have it
\font\fivemib=cmmib10 at 5pt  % =cmmib5 % if you have it
\font\tenbsy=cmbsy10
\font\sevenbsy=cmbsy10 at 7pt % =cmbsy7 % if you have it
\font\fivebsy=cmbsy10 at 5pt  % =cmbsy5 % if you have it
\def\BMfont{\textfont0\tenbf \scriptfont0\sevenbf
                              \scriptscriptfont0\fivebf
            \textfont1\tenmib \scriptfont1\sevenmib
                               \scriptscriptfont1\fivemib
            \textfont2\tenbsy \scriptfont2\sevenbsy
                               \scriptscriptfont2\fivebsy}
\def\rlx{\relax\leavevmode}
 % Guess what this is for...
\def\BM#1{\rlx\ifmmode\mathchoice
                      {\hbox{$\BMfont#1$}}
                      {\hbox{$\BMfont#1$}}
                      {\hbox{$\scriptstyle\BMfont#1$}}
                      {\hbox{$\scriptscriptstyle\BMfont#1$}}
                 \else{$\BMfont#1$}\fi}

\font\tenmib=cmmib10
\font\sevenmib=cmmib10 at 7pt % =cmmib7 % if you have it
\font\fivemib=cmmib10 at 5pt  % =cmmib5 % if you have it
\font\tenbsy=cmbsy10
\font\sevenbsy=cmbsy10 at 7pt % =cmbsy7 % if you have it
\font\fivebsy=cmbsy10 at 5pt  % =cmbsy5 % if you have it
\def\BMfont{\textfont0\tenbf \scriptfont0\sevenbf
                              \scriptscriptfont0\fivebf
            \textfont1\tenmib \scriptfont1\sevenmib
                               \scriptscriptfont1\fivemib
            \textfont2\tenbsy \scriptfont2\sevenbsy
                               \scriptscriptfont2\fivebsy}
\def\BM#1{\rlx\ifmmode\mathchoice
                      {\hbox{$\BMfont#1$}}
                      {\hbox{$\BMfont#1$}}
                      {\hbox{$\scriptstyle\BMfont#1$}}
                      {\hbox{$\scriptscriptstyle\BMfont#1$}}
                 \else{$\BMfont#1$}\fi}

\def\inbar{\vrule height1.5ex width.4pt depth0pt}
\def\sinbar{\vrule height1ex width.35pt depth0pt}
\def\ssinbar{\vrule height.7ex width.3pt depth0pt}
\font\cmss=cmss10
\font\cmsss=cmss10 at 7pt
\def\ZZ{\rlx\leavevmode
             \ifmmode\mathchoice
                    {\hbox{\cmss Z\kern-.4em Z}}
                    {\hbox{\cmss Z\kern-.4em Z}}
                    {\lower.9pt\hbox{\cmsss Z\kern-.36em Z}}
                    {\lower1.2pt\hbox{\cmsss Z\kern-.36em Z}}
               \else{\cmss Z\kern-.4em Z}\fi}
\def\Ik{\rlx{\rm I\kern-.18em k}}  % Yes, I know. This ain't capital.
\def\IC{\rlx\leavevmode
             \ifmmode\mathchoice
                    {\hbox{\kern.33em\inbar\kern-.3em{\rm C}}}
                    {\hbox{\kern.33em\inbar\kern-.3em{\rm C}}}
                    {\hbox{\kern.28em\sinbar\kern-.25em{\rm C}}}
                    {\hbox{\kern.25em\ssinbar\kern-.22em{\rm C}}}
             \else{\hbox{\kern.3em\inbar\kern-.3em{\rm C}}}\fi}
\def\IP{\rlx{\rm I\kern-.18em P}}
\def\IR{\rlx{\rm I\kern-.18em R}}
\def\IN{\rlx{\rm I\kern-.20em N}}
\def\Ione{\rlx{\rm 1\kern-2.7pt l}}
% ----------------- End of Additional Fonts -----------------

% ----------------- harvmactest.tex -------------------------
%% site dependent options:
%% \unredoffs and \redoffs define horizontal and vertical offsets
%% respectively for unreduced and reduced modes. \speclscape defines
%% the \special{} call that sets printer to landscape (sideways) mode.
%% from standard set below, leave uncommented as appropriate or redefine
%
%%% next 400dpi
%\def\unredoffs{} \def\redoffs{\voffset=-.31truein\hoffset=-.48truein}
%\def\speclscape{\special{landscape}}
%
%%% apple lw
\def\unredoffs{} \def\redoffs{\voffset=-.31truein\hoffset=-.59truein}
\def\speclscape{\special{ps: landscape}}

\newbox\leftpage \newdimen\fullhsize \newdimen\hstitle \newdimen\hsbody
\tolerance=1000\hfuzz=2pt\def\fontflag{cm}
\catcode`\@=11 % This allows us to modify PLAIN macros.
% We need next two \doit commands to avoid the repeated questions of reduction.
\doit0
{
\def\bigans{b }
\message{ big or little (b/l)? }\read-1 to\answ
\ifx\answ\bigans\message{(This will come out unreduced.}
}
%\magnification=1200\unredoffs\baselineskip=16pt plus 2pt minus 1pt
\hsbody=\hsize \hstitle=\hsize %take default values for unreduced format
\doit0{
\else\message{(This will be reduced.} \let\l@r=L
%\magnification=1000\baselineskip=16pt plus 2pt minus 1pt \vsize=7truein
\redoffs \hstitle=8truein\hsbody=4.75truein\fullhsize=10truein\hsize=\hsbody
\output={\ifnum\pageno=0 %%% This is the HUTP version
  \shipout\vbox{\speclscape{\hsize\fullhsize\makeheadline}
    \hbox to \fullhsize{\hfill\pagebody\hfill}}\advancepageno
  \else
  \almostshipout{\leftline{\vbox{\pagebody\makefootline}}}\advancepageno
  \fi}
}
\def\almostshipout#1{\if L\l@r \count1=1 \message{[\the\count0.\the\count1]}
      \global\setbox\leftpage=#1 \global\let\l@r=R
 \else \count1=2
  \shipout\vbox{\speclscape{\hsize\fullhsize\makeheadline}
      \hbox to\fullhsize{\box\leftpage\hfil#1}}  \global\let\l@r=L\fi}
\fi
%---------------------------------------------------------------------
%       use \nolabels to get rid of eqn, ref, and fig labels in draft mode
\def\nolabels{\def\wrlabeL##1{}\def\eqlabeL##1{}\def\reflabeL##1{}}
\def\writelabels{\def\wrlabeL##1{\leavevmode\vadjust{\rlap{\smash%
{\line{{\escapechar=` \hfill\rlap{\sevenrm\hskip.03in\string##1}}}}}}}%
\def\eqlabeL##1{{\escapechar-1\rlap{\sevenrm\hskip.05in\string##1}}}%
\def\reflabeL##1{\noexpand\llap{\noexpand\sevenrm\string\string\string##1}}}
\nolabels
%
% tagged sec numbers
\global\newcount\secno \global\secno=0
\global\newcount\meqno \global\meqno=1
\def\newsec#1{\global\advance\secno by1\message{(\the\secno. #1)}
%\ifx\answ\bigans \vfill\eject \else \bigbreak\bigskip \fi  %if desired
\global\subsecno=0\eqnres@t\noindent{\bf\the\secno. #1}
\writetoca{{\secsym} {#1}}\par\nobreak\medskip\nobreak}
\def\eqnres@t{\xdef\secsym{\the\secno.}\global\meqno=1\bigbreak\bigskip}
\def\sequentialequations{\def\eqnres@t{\bigbreak}}\xdef\secsym{}
\global\newcount\subsecno \global\subsecno=0
\def\subsec#1{\global\advance\subsecno by1\message{(\secsym\the\subsecno. #1)}
\ifnum\lastpenalty>9000\else\bigbreak\fi
\noindent{\it\secsym\the\subsecno. #1}\writetoca{\string\quad
{\secsym\the\subsecno.} {#1}}\par\nobreak\medskip\nobreak}
\def\appendix#1#2{\global\meqno=1\global\subsecno=0\xdef\secsym{\hbox{#1.}}
\bigbreak\bigskip\noindent{\bf Appendix #1. #2}\message{(#1. #2)}
\writetoca{Appendix {#1.} {#2}}\par\nobreak\medskip\nobreak}
%
%       \eqn\label{a+b=c}	gives displayed equation, numbered
%				consecutively within sections.
%     \eqnn and \eqna define labels in advance (of eqalign?)
%
\def\eqnn#1{\xdef #1{(\secsym\the\meqno)}\writedef{#1\leftbracket#1}%
\global\advance\meqno by1\wrlabeL#1}
\def\eqna#1{\xdef #1##1{\hbox{$(\secsym\the\meqno##1)$}}
\writedef{#1\numbersign1\leftbracket#1{\numbersign1}}%
\global\advance\meqno by1\wrlabeL{#1$\{\}$}}
\def\eqn#1#2{\xdef #1{(\secsym\the\meqno)}\writedef{#1\leftbracket#1}%
\global\advance\meqno by1$$#2\eqno#1\eqlabeL#1$$}
%
%			 footnotes
\newskip\footskip\footskip14pt plus 1pt minus 1pt %sets footnote baselineskip
\def\footnotefont{\ninepoint}\def\f@t#1{\footnotefont #1\@foot}
\def\f@@t{\baselineskip\footskip\bgroup\footnotefont\aftergroup\@foot\let\next}
\setbox\strutbox=\hbox{\vrule height9.5pt depth4.5pt width0pt}
\global\newcount\ftno \global\ftno=0
\def\foot{\global\advance\ftno by1\footnote{$^{\the\ftno}$}}
%
%say \footend to put footnotes at end
%will cause problems if \ref used inside \foot, instead use \nref before
\newwrite\ftfile
\def\footend{\def\foot{\global\advance\ftno by1\chardef\wfile=\ftfile
$^{\the\ftno}$\ifnum\ftno=1\immediate\openout\ftfile=foots.tmp\fi%
\immediate\write\ftfile{\noexpand\smallskip%
\noexpand\item{f\the\ftno:\ }\pctsign}\findarg}%
\def\footatend{\vfill\eject\immediate\closeout\ftfile{\parindent=20pt
\centerline{\bf Footnotes}\nobreak\bigskip\input foots.tmp }}}
\def\footatend{}
%
%     \ref\label{text}
% generates a number, assigns it to \label, generates an entry.
% To list the refs on a separate page,  \listrefs
%
\global\newcount\refno \global\refno=1
\newwrite\rfile
%% We have tampered after #1 in \items which was originally \item and also
%% the argument of \xdef without [ ].  Also \\ after \items{#1}.
%
\def\ref{[\the\refno]\nref}%
\def\nref#1{\xdef#1{[\the\refno]}\writedef{#1\leftbracket#1}%
\ifnum\refno=1\immediate\openout\rfile=refs.tmp\fi%
\global\advance\refno by1\chardef\wfile=\rfile\immediate%
\write\rfile{\noexpand\Item{#1}\reflabeL{#1\hskip.31in}\pctsign}%
\findarg\hskip10.0pt}%
%	horrible hack to sidestep tex \write limitation
\def\findarg#1#{\begingroup\obeylines\newlinechar=`\^^M\pass@rg}
{\obeylines\gdef\pass@rg#1{\writ@line\relax #1^^M\hbox{}^^M}%
\gdef\writ@line#1^^M{\expandafter\toks0\expandafter{\striprel@x #1}%
\edef\next{\the\toks0}\ifx\next\em@rk\let\next=\endgroup\else\ifx\next\empty%
\else\immediate\write\wfile{\the\toks0}\fi\let\next=\writ@line\fi\next\relax}}
\def\striprel@x#1{} \def\em@rk{\hbox{}}
\def\lref{\begingroup\obeylines\lr@f}
\def\lr@f#1#2{\gdef#1{\ref#1{#2}}\endgroup\unskip}
\def\semi{;\hfil\break}
\def\addref#1{\immediate\write\rfile{\noexpand\item{}#1}} %now unnecessary
\def\listrefs{\footatend\vfill\supereject\immediate\closeout\rfile\writestoppt
\baselineskip=14pt\centerline{{\bf References}}\bigskip{\frenchspacing%
\parindent=20pt\escapechar=` \input refs.tmp\vfill\eject}\nonfrenchspacing}
%
% The following is the revision of \listrefs to put the list in the same page.
\def\listrefsr{\immediate\closeout\rfile\writestoppt
\baselineskip=14pt\centerline{{\bf References}}\bigskip{\frenchspacing%
\parindent=20pt\escapechar=` \input refs.tmp\vfill\eject}\nonfrenchspacing}
\def\startrefs#1{\immediate\openout\rfile=refs.tmp\refno=#1}
\def\xref{\expandafter\xr@f}\def\xr@f[#1]{#1}
\def\refs#1{\count255=1[\r@fs #1{\hbox{}}]}
\def\r@fs#1{\ifx\und@fined#1\message{reflabel \string#1 is undefined.}%
\nref#1{need to supply reference \string#1.}\fi%
\vphantom{\hphantom{#1}}\edef\next{#1}\ifx\next\em@rk\def\next{}%
\else\ifx\next#1\ifodd\count255\relax\xref#1\count255=0\fi%
\else#1\count255=1\fi\let\next=\r@fs\fi\next}
\def\figures{\centerline{{\bf Figure Captions}}\medskip\parindent=40pt%
\def\fig##1##2{\medskip\item{Fig.~##1.  }##2}}
%
% this is ugly, but moore insists
\newwrite\ffile\global\newcount\figno \global\figno=1
\def\fig{fig.~\the\figno\nfig}
\def\nfig#1{\xdef#1{fig.~\the\figno}%
\writedef{#1\leftbracket fig.\noexpand~\the\figno}%
\ifnum\figno=1\immediate\openout\ffile=figs.tmp\fi\chardef\wfile=\ffile%
\immediate\write\ffile{\noexpand\medskip\noexpand\item{Fig.\ \the\figno. }
\reflabeL{#1\hskip.55in}\pctsign}\global\advance\figno by1\findarg}
\def\listfigs{\vfill\eject\immediate\closeout\ffile{\parindent40pt
\baselineskip14pt\centerline{{\bf Figure Captions}}\nobreak\medskip
\escapechar=` \input figs.tmp\vfill\eject}}
\def\xfig{\expandafter\xf@g}\def\xf@g fig.\penalty\@M\ {}
\def\figs#1{figs.~\f@gs #1{\hbox{}}}
\def\f@gs#1{\edef\next{#1}\ifx\next\em@rk\def\next{}\else
\ifx\next#1\xfig #1\else#1\fi\let\next=\f@gs\fi\next}
\newwrite\lfile
{\escapechar-1\xdef\pctsign{\string\%}\xdef\leftbracket{\string\{}
\xdef\rightbracket{\string\}}\xdef\numbersign{\string\#}}
\def\writedefs{\immediate\openout\lfile=labeldefs.tmp \def\writedef##1{%
\immediate\write\lfile{\string\def\string##1\rightbracket}}}
\def\writestop{\def\writestoppt{\immediate\write\lfile{\string\pageno%
\the\pageno\string\startrefs\leftbracket\the\refno\rightbracket%
\string\def\string\secsym\leftbracket\secsym\rightbracket%
\string\secno\the\secno\string\meqno\the\meqno}\immediate\closeout\lfile}}
\def\writestoppt{}\def\writedef#1{}
\def\seclab#1{\xdef #1{\the\secno}\writedef{#1\leftbracket#1}\wrlabeL{#1=#1}}
\def\subseclab#1{\xdef #1{\secsym\the\subsecno}%
\writedef{#1\leftbracket#1}\wrlabeL{#1=#1}}
\newwrite\tfile \def\writetoca#1{}
\def\leaderfill{\leaders\hbox to 1em{\hss.\hss}\hfill}
%	use this to write file with table of contents
\def\writetoc{\immediate\openout\tfile=toc.tmp
   \def\writetoca##1{{\edef\next{\write\tfile{\noindent ##1
   \string\leaderfill {\noexpand\number\pageno} \par}}\next}}}
%       and this lists table of contents on second pass
\def\listtoc{\centerline{\bf Contents}\nobreak\medskip{\baselineskip=12pt
 \parskip=0pt\catcode`\@=11 \input toc.tex \catcode`\@=12 \bigbreak\bigskip}}
\catcode`\@=12 % at signs are no longer letters
%
% --------------- end of harvmac.tex --------------------

\def\items#1{\\ \item{[#1]}}

\def\dfno{~$D=4,\,N=1\,$~}
\def\dtnt{~$D=3,\,N=2\,$~}

\def\framing#1{\doit{#1}
{\framingfonts{#1}
\border\headpic
}}

\framing{0}

{}~~~
\vskip 0.07in

\doit1{
{\hbox to\hsize{
October 1995 \hfill UMDEPP 96$\,$--$\,$42}}
%{\hbox to\hsize{~~~~~ ~~~~~~\hfill (Revised Version)}}
}
\par

%\hfill {(Revised Version)}\\

\hsize=6.5in
\textwidth=4.5in
\oddsidemargin=0.07in

\begin{center}
\vglue 0.1in

{\large\bf Conical ~Singularities ~in ~Three ~or ~Four~-~Dimensions} \\
{\large\bf and ~Supersymmetry ~Breaking}$\,$\footnote{This work is
supported in part by NSF grant \# PHY-93-41926
and by DOE grant \# DE-FG02-94ER40854.
}
\\[.1in]

\baselineskip 10pt

\vskip 0.28in

\doit0{S.~James GATES, Jr.\footnote{E-mail: gates@umdhep.umd.edu}~ and~
Hitoshi NISHINO\footnote{Also at  Department of Physics and Astronomy,
Howard University, Washington, D.C. 20059, USA.
E-mail: nishino@umdhep.umd.edu.} \\[.25in]
}

\doit0{
Hitoshi ~N{\small ISHINO}\footnote{E-mail: nishino@umdhep.umd.edu.}
\\[.25in]
{\it Department of Physics and Astronomy} \\[.015in]
{\it Howard University} \\[.015in]
{\it Washington, D.C. 20059, USA} \\[.18in]
and \\[.18in]
{\it Department of Physics} \\ [.015in]
{\it University of Maryland at College Park}\\ [.015in]
{\it College Park, MD 20742-4111, USA} \\[.18in]
}

\doit1{
Hitoshi ~N{\small ISHINO}\footnote{E-mail: nishino@umdhep.umd.edu.}
\\[.25in]
{\it Department of Physics} \\ [.015in]
{\it University of Maryland at College Park}\\ [.015in]
{\it College Park, MD 20742-4111, USA} \\[.12in]
and \\[.12in]
{\it Department of Physics and Astronomy} \\[.015in]
{\it Howard University} \\[.015in]
{\it Washington, D.C. 20059, USA} \\[.18in]
}

\vskip 2.0in

{\bf Abstract} \\[.1in]

\end{center}

\begin{quotation}

{}~~~We present mechanisms for generating conical singularities both in three
and four-dimensions in the systems with copies of scalar or chiral
multiplets coupled to $~N=2$~ or $~N=1$~ supergravity.
Our mechanisms are useful for supersymmetry breaking, maintaining the
zero cosmological constants in three and four-dimensions.  A strong
coupling duality connecting these two dimensionalities is also
studied.

\endtitle
\vfill\eject

\oddsidemargin=0.03in
\evensidemargin=0.01in
\hsize=6.5in
\textwidth=6.5in
\baselineskip 16.5pt

\centerline{\bf 1.~~Introduction}

It is some years ago that E.~Witten proposed an original scenario of
supersymmetry breaking \ref\witten{E.~Witten, {\it ``Is
Supersymmetry Really Broken?''}, talk given at the {\it
``Santa Fe Institute Meeting on String Theory''}; \ijmp{10}{95}{1247}.}
maintaining the zero cosmological constant in three-dimensional space-time
($D=3$) with conical singularities \ref\djt{S.~Deser, R.~Jackiw and
G.~'t Hooft, \ap{152}{84}{220}.}.  Such a conical space-time has a deficit
angle which lifts the degeneracy between bosons and fermions in a
supermultiplet yielding supersymmetry  breaking.  An explicit mechanism
realizing this mechanism has also been presented by K.~Becker {\it et
al.}~\ref\bbs{K.~Becker, M.~Becker and A.~Strominger, \pr{51}{95}{6603}.}
based on $~D=3,~N=2$~ supergravity coupled to a vector multiplet and a
charged scalar multiplet, with a Nielsen-Olesen vortex soliton solution
\ref\no{H.B.~Nielsen and P.~Olesen, \np{61}{73}{45}.}.  However, there has
been no indication  that a direct analog of this mechanism works in
four-dimensions \bbs.

Recently there has been a nice scenario
that the mystery about the vanishing
cosmological constant in $~D=4$~ even after the supersymmetry breaking can
be solved by the strong coupling duality \ref\wittenduality{E.~Witten,
{\it ``Some Comments on String Dynamics''},
preprint, IASSNS-HEP-95-63 (July 1995) hep-th/9507121; {\it ``Strong
Coupling and the Cosmological Constant''}, preprint, IASSNS-HEP-95-51
(June 1995) hep-th/9506101.} between the $~D=3$~ theory with
broken supersymmetry with conical singularity and the $~D=4$~ theory
regarded as its strong coupling limit $~\l\rightarrow\infty$.  In
particular, the $~D=4$~ Poincar\'e invariance with a zero cosmological
constant is recovered, because the radius $~r$~ of the circle in
$~\IR^3\otimes S^1$~ grows: $~r\rightarrow\infty$~ as $~\l\rightarrow\infty$~
\wittenduality.

In this paper we present an explicit model of generating such conical
singularity for $~D=3,\,N=2$~ supergravity different from that in \bbs, by
taking a limit of a particular mass to infinity, which is applicable also to
$~D=4,\,N=1$~ supergravity resulting in a desirable supersymmetry breaking
maintaining the vanishing cosmological constant.  We also see if the recent
scenario of the strong coupling duality \wittenduality\ works between the
two systems we deal with.

\bigskip\bigskip\smallskip

%\newpage

\centerline{\bf 2.~~$D=3,~N=2$~ Supergravity Coupled to Scalar Multiplets}

Before presenting our mechanism of conical singularity, we first fix the
invariant lagrangian of the $~D=3,~N=2$~ supergravity.  Our field contents
are the supergravity multiplet $~(e\du\m m, \psi\du\m i, B_\m, \chi^i,
\varphi)$~ with the indices $~{\scst i,~j,~\cdots~=~1,~2}$~ for the
doublets under the global $~SO(2)$~ of the $~N=2$~ supersymmetry, and $~n$~
copies of scalar multiplets $~(A_a, B_a, \l_a^i)~{\scst
(a,~b,~\cdots~=~1,~\cdots,~n)}$.
%%%%%%%%%%%%%%%%%%%%%%%%%%%%%%%%%%%%%%%%%%%%%%%%%%%%%%%%%%%%%%%%%%%%%%%%%%%
% To Mr./Ms. Type-Setter in charge, please do NOT delete the \scst
% command (abbreviation of \scriptstyle)
% in the above line for print-setting.  This command is definitely
% needed to distinguish these ``subscripts'' from the constant $a$
% coming later in (3.7).  Otherwise the results will be very confusing.
%%%%%%%%%%%%%%%%%%%%%%%%%%%%%%%%%%%%%%%%%%%%%%%%%%%%%%%%%%%%%%%%%%%%%%%%%%%
The indices $~{\scst\m,~\n,~\cdots~=~0,~1,~2}$~ are for curved coordinates,
while $~{\scst m,~n,~\cdots~=~(0),~(1),~(2)}$~ are local Lorentz indices
with the signature ~$(+,-,-)$.  We do not gauge the global $~SO(2)$~
symmetry in order not to create a potential with a generally non-zero
cosmological constant.

Our invariant lagrangian for $~N=2$~ supergravity is\footnotew{To our
knowledge,  the lagrangians (2.1) - (2.3) have never been presented in the
past.  These lagrangians are fixed up to fermionic quartic terms.}   $$
\eqalign{e^{-1}\Lag_{\rm SG} = \, & - \fracm 1 4 R + \fracm 1 2
e^{-1}\e^{\m\n\r} \left( \Bar\psi\du\m i D_\n(\Hat\o)\psi\du\r i \right)
-\fracm 1 4 e^{2 b \varphi} G_{\m\n}^2 + \fracm i 2 \left( \Bar\chi{\,}^i
\g^\m D_\m(\Hat\o) \chi^i \right) \cr
& + \half (\partial_\m\varphi)^2 - \fracm i {2\sqrt2} \e^{i j} \left(
\Bar\psi{}\du\m i \g^{\r\s} \g^\m \chi^j \right) e^{b\varphi} G_{\r\s} -
\fracm 1 {\sqrt2} \left( \Bar\psi{} \du\m i \g^\n\g^\m \chi^i \right)
\partial_\n \varphi \cr
& - \fracm 1{2b} \e^{i j} e^{b\varphi} \left( \Bar\psi{}\du\m i \psi\du\n j
\right) G^{\m\n} + \fracm 1 4 \left( \fracm 1 b - b \right) \e^{i j}
e^{b\varphi} \left( \Bar\chi{\,}^i \g^{\m\n} \chi^j \right) G_{\m\n} ~~.
\cr }
\eqno(2.1) $$

\vfill\eject

\noindent The constant $~b$~ is an arbitrary non-zero real number.  The kinetic
lagrangian for the scalar multiplets is
$$ \eqalign{e^{-1} \Lag_{\rm SM} = \sum_a \bigg[ \, & + \half e^{-b\varphi}
(\partial_\m A_a)^2 + \half e^{- b\varphi} (\partial_\m B_a)^2 + \fracm i 2
\left( \Bar\l{}_a^i \g^\m D_\m(\Hat\omega) \l_a^i \right) \cr
& - \fracm1{\sqrt2} \left( \Bar\psi{} \du\m i \g^\n\g^\m \l_a^i \right)
e^{-b\varphi/2} \partial_\n A_a - \fracm 1{\sqrt2} \e^{i j}
\left( \Bar\psi\du\m i
\g^\n\g^\m \l_a^j \right) e^{-b\varphi/2} \partial_\n B_a \cr
& + b e^{-1}
\e^{\m\n\r} B_\m (\partial_\n A_a) (\partial_\r B_a)
- \fracm 1 8 \left( b - \fracm 2 b  \right) \e^{i j} \left(
\Bar\l{}_a^i \g^{\m\n} \l_a^j \right) e^{b\varphi} G_{\m\n} \cr
& + \fracm i 2 b \left(
\Bar\chi{\,}^i
\g^\m \l_a^i \right) e^{-b\varphi/2} \partial_\m A_a  +\fracm i 2 b
\e^{i j} \left( \Bar\chi{\,}^i \g^\m \l_a^j \right) e^{- b \varphi/2}
\partial_\m  B_a
\, \bigg]  ~~. \cr   }
\eqno(2.2)  $$
The invariant mass terms are\footnotew{We have not been
successful for finding out a more general potential terms.}
$$\eqalign{e^{-1} \Lag_m =  \sum_a \bigg[ & - \half m_a^2 e^{-b\varphi}
\left( A_a^2+B_a^2 \right) + \half m_a \left(\Bar\l{}_a^i\l_a^i  \right) \cr
& + \fracm i {\sqrt 2} m_a \left( \Bar\psi{}\du\m i \g^\m \l_a^i \right)
e^{-b\varphi/2} A_a + \fracm i {\sqrt2} m_a \e^{i j} \left( \Bar\psi{}\du
\m i
\g^\m \l_a^j \right) e^{-b\varphi/2} B_a \cr
& - \half b \, m_a \left( \Bar\chi{\,}^i\l_a^i \right) e^{-b\varphi/2} A_a
- \half b \, m_a \e^{i j}\left( \Bar\chi{\,}^i\l_a^j \right) e^{-b\varphi/2}
B_a\bigg] ~~. \cr}
\eqno(2.3) $$

The total lagrangian $~\Lag_{\rm total} \equiv \Lag_{\rm SG} +
\Lag_{\rm SM}+\Lag_m $~ up to quartic fermion terms is invariant  under
the  supertranslation rules up to bilinear fermion terms
$$\li{& \d e\du\m m = - i\left(\Bar\e{\,}^i \g^m \psi\du\m i \right)
{}~~,  &(2.4a) \cr   &\d\psi\du\m i = D_\m (\Hat\omega) \e{\,}^i + \fracm 1
{2b} e^{-1} \e\du\m{\r\s} \e^{i j} \e^j e^{b\varphi} \Hat G_{\r\s} ~~,
&(2.4b) \cr
&\d B_\m = \fracm i {\sqrt2}
e^{- b\varphi} \e^{i j} \left( \Bar\e{\,}^i \g_\m\chi^j \right)  -
\fracm 1 b
e^{-b\varphi} \e^{i j} \left(\Bar\e{\,}^i\psi\du\m j \right) ~~,
&(2.4c) \cr
&\d\chi^i = - \fracm 1 {2\sqrt2} \e^{i j} \g^{\m\n} \e^j e^{b \varphi} \Hat
G_{\m\n} - \fracm i {\sqrt 2} \g^\m\e^i \Hat D_\m \varphi ~~,
&(2.4d) \cr
&\d\varphi = + \fracm 1{\sqrt2} \left( \Bar\e{\,}^i \chi^i \right) ~~,
&(2.4e) \cr
& ~~~~~ \cr
&\d A_a = \fracm1{\sqrt2} e^{b\varphi/2} \left( \Bar\e{\,}^i\l_a^i
\right) ~~,
&(2.5a) \cr
&\d B_a = \fracm 1{\sqrt 2} \e^{i j} e^{b\varphi/2} \left( \Bar\e{\,}^i
\l_a^j \right) ~~,
&(2.5b) \cr
&\d\l_a^i = -\fracm i {\sqrt2} \left( \g^\m \e^i\right)
e^{-b\varphi/2} \partial_\m A_a + \fracm i {\sqrt2} \e^{i j} \left( \g^\m
\e^j\right) e^{-b\varphi/2} \partial_\m B_a  \cr
&~~~~~ ~~~ + \frac1 {\sqrt2} m_a \e^i e^{-b\varphi/2} A_a -
\frac1 {\sqrt2} m_a \e^{i j} \e^j e^{-b\varphi/2} B_a~~.
&(2.5c) \cr}  $$
As usual \ref\salam{A.~Salam and E.~Sezgin, {\it ``Supergravities in
Diverse Dimensions''}, Vols.~I and II (Elsevier Science Publishers,
B.V.~and World Scientific Publishers, Co.~Ltd., 1989).}, this system has a
global symmetry associated with the dilaton:
$$ \varphi \rightarrow \varphi + c~~,
{}~~~~B_\m \rightarrow e^{-b c} B_\m~~, ~~~~A_a\rightarrow e^{b c/2} A_a
{}~~, ~~~~  B_a\rightarrow e^{b c/2} B_a~~,
\eqno(2.6) $$
with an arbitrary non-zero real constant parameter $~c$.
Additionally, these lagrangians are scaling:
$$\Lag_{\rm SG} \rightarrow e^{a c} \Lag_{\rm SG} ~~, ~~~~
\Lag_{\rm SM} \rightarrow e^{a c} \Lag_{\rm SM} ~~, ~~~~
\Lag_m \rightarrow e^{a c} \Lag_m~~,
\eqno(2.7a) $$
under
$$\eqalign{ &\varphi \rightarrow \varphi + c~~, \cr
&(e\du\m m , \psi_\m , B_\m , \chi) \rightarrow (e^{a c} e\du\m m ,
e^{a c/2} \psi_\m, e^{(a-b) c} B_\m, e^{-a c/2} \chi) ~~, \cr
& (A_a , B_a, \l) \rightarrow (e^{b c/2}  A_a, e^{b c/2}  B_a,
e^{-a c/2} \l) ~~, ~~~~ m \rightarrow e^{-a c} m ~~, \cr }
\eqno(2.7b) $$
where $~c$~ is an arbitrary constant parameter, while $~a$~ is a real
constant.  (For simplicity we can choose $~a=b$.)

This system has a zero cosmological constant, unless the global
$~SO(2)$~ symmetry is gauged by minimal couplings with the
$~B_\m\-$field.  We can further introduce a supersymmetric
Chern-Simons lagrangian with a $~B\wedge G\-$term with a non-zero potential
term,  which we skip in this paper due to its irrelevance to our present
purpose.

There are some remarks in order.  The $~D=3,\,N=2$~ system presented
here is related to eqs.~(B.1) - (B.4) given in ref.~\ref\ng{H.~Nishino and
S.J.~Gates, \ijmp{8}{93}{3371}.} in the way that the irreducible
supergravity multiplet  $~(e\du\m m, \psi_\m)$~ and a vector multiplet
$~(A_\m, \l, S)$~ in \ng\ are combined to form a reducible supergravity
multiplet (2.4) with  $~(A_\m, \l,S)~$ replaced by
$~(B_\m,\chi,\varphi)$.  In particular, the scalar field $~S$~ in \ng\
is now the dilaton in our system with non-polynomial couplings.
This can be understood as follows.  A simple computation reveals that the
kinetic lagrangian for the vector multiplet in \ng\ generates what is called
``improvement term''$\,\approx S^2 R(\omega)$~ in the lagrangian.
In fact, the supertranslation of the auxiliary field $~D$~ for the
vector multiplet \ng\ indicates the gaugino field equation with the term
$~\approx\left(\g^{\m\n}D_{\[\m}\psi_{\n\]}\right)S$~ as in (B.4) in \ng,
implying the existence of the term $~\approx\left(\Bar\l\g^{\m\n}D_{\[\m}
\psi_{\n\]}\right)S$~ in the lagrangian, which in turn necessitates the
improvement term $~\approx S^2 R(\omega)$~ above.  This signals the mixture
between the dreibein and the
$~S\-$field.  To have a {\it canonical} system free from such a mixture, we
have  to rescale the dreibein in such a way that this scalar appears in the
lagrangian with non-polynomial couplings.  The simplest principle is to
interpret this scalar as the dilaton with exponential couplings, as we have
done here.  Accordingly, there arise cross terms in the
supertranslation rules between the original supergravity and vector
multiplets, such as the $~G\-$term in (2.4b).   It is for this reason that
we are dealing with our enlarged supergravity  multiplet in this
paper.\footnotew{From these considerations, we do not completely agree with
the polynomial couplings of the $~N$-field in ref.~\bbs.  We claim that
either non-polynomial couplings of the $~N$-field, or the improvement term
$~\approx N^2 R(\omega)$~ should 	`arise in the lagrangian.  This is also
reasonable from the viewpoint of duality \wittenduality, connecting $~D=4$~ and
$~D=3$~ supergravities, because the dilaton in the former will disappear in
the latter under the strong coupling duality.  (Cf.~(5.3))}

\bigskip\bigskip\bigskip

%\newpage

\centerline{\bf 3.~~Conical Singularity in $~D=3$}

We now present our mechanism generating conical singularities in $~D=3$.
We start with the bosonic field equations from $~\Lag_{\rm total}$:

\vfill\eject

$$\li{ & R_{\m\n} = - 2e^{2b \varphi} G_{\m\r}G\du\n\r + g_{\m\n}
e^{2b\varphi} G_{\r\s}^2 + 2(\partial_\m\varphi)(\partial_\n\varphi)  \cr
& ~~~~~ ~~~~~  + \sum_a\bigg[ 2 e^{-b\varphi} (\partial_\m A_a)(\partial_\n
A_a) + 2 e^{-b\varphi} (\partial_\m B_a)(\partial_\n B_a)
- 2m_a^2 g_{\m\n} e^{-b\varphi} (A_a^2 + B_a^2) \bigg] ~,\hbox{~~~~}
&(3.1) \cr
& D_\m^2 \varphi = - \half b e^{2b\varphi} G_{\m\n}^2  \cr
& ~~~~~ ~~~~~ + \sum_a \bigg[ - \fracm b 2 e^{-b\varphi}
(\partial_\m A_a)^2 - \fracm b 2 e^{-b\varphi} (\partial_\m B_a)^2
 + \fracm b 2 m_a^2 e^{-b\varphi} (A_a^2 + B_a^2) \bigg] ~~, {~~~~}
&(3.2) \cr
& \partial_\n \left(ee^{2b\varphi} G^{\m\n}\right)  = - b \e^{\m\n\r}\sum_a
(\partial_\n A_a) (\partial_\r B_a) ~~,
&(3.3) \cr
&\partial_\m(ee^{-b\varphi} g^{\m\n} \partial_\n A_a) - \fracm b 2
\e^{\m\n\r} G_{\m\n} \partial_\r B_a + m_a ^2 e e^{-b\varphi} A_a
= 0 ~~,
&(3.4) \cr
& \partial_\m(ee^{-b\varphi} g^{\m\n} \partial_\n B_a) + \fracm b 2
\e^{\m\n\r} G_{\m\n} \partial_\r A_a + m_a ^2 e e^{-b\varphi} B_a = 0~~.
&(3.5) \cr} $$

Our ansatz for the dreibein is
$$(e\du\m m) = \pmatrix{1 & 0 & 0 \cr 0 & e^{\s} & 0 \cr 0 & 0 & e^{\s}
\cr} ~~,
\eqno(3.6) $$
for the coordinates $~(x^0,x^1,x^2)=(t,x,y)$.  If we have the solution
$$ \s = a \ln r ~~, ~~~~ r \equiv {\sqrt {x^2+y^2}}~~~~~~
(a:~\hbox{const.})~~,
\eqno(3.7)$$
for a non-integer $~a$, we have a conical singularity
\djt.  For simplicity we also require the backgrounds
$$ \varphi = 0 ~~, ~~~~G_{\m\n} = 0 ~~, ~~~~ B_1 = 0 ~~, ~~~~ A_a = B_a =0
{}~~~~~~({\scst a~\ge ~2})~~.
\eqno(3.8) $$
Accordingly, eqs.~(3.4) for $~{\scst a~\ge ~2}$~ and (3.5) for
$~{\scst a~\ge ~1}$~ are trivially satisfied.  From now on, we use $~M\equiv
m_1$~ distinguished  from other masses.

We next solve eq.~(3.4) for $~{\scst a~=~1}$~ under the ansatz (3.8):
$$\partial_x ^2 A_1 + \partial_y ^2 A_1 - M^2 e^{2\s} A_1
= A_1''(r) + \fracm 1 r A_1'(r) - M^2 r^{2a} A_1(r)  = 0 ~~,
\eqno(3.9) $$
where the primes are derivatives by the variables in the
parentheses.  A new variable
$$ \zeta \equiv \fracm M {a+1} r^{a+1} ~~, ~~~~ r = \left(
\fracm{a+1} M \right)^{1/(a+1)} \zeta^{1/(a+1)} ~~,
\eqno(3.10) $$
simplifies (3.9) as \ref\gr{{\it See e.g.}, I.S.~Gradshteyn and
Ryzhik, {\it ``Table of Integrals, Series and Products''}
(Academic Press, 1980).}
$$ A_1''(\z) + \fracm 1 \z A_1'(\z) - A_1(\z) = 0 ~~,
\eqno(3.11) $$
which can be easily solved by
$$ A_1 = \a K_0(\z) = \a K_0 \left( \fracm M {a+1} r^{a+1} \right)
{}~~~~(\a:~\hbox{const.})~~,
\eqno(3.12) $$
as long as
$$ a > -1 ~~.
\eqno(3.13) $$
Here $~K_0(z)$~ is a ``modified Bessel function'' related to the Hankel
functions (Bessel functions of the third kind) \gr\ as
$$ K_\n (z) \equiv \fracm {\pi i} 2 \,  e^{\n\pi i/2} \, H_n^{(1)} (i z)~~.
\eqno(3.14) $$
Eq.~(3.13) is needed for the damping behaviour at $~r\rightarrow\infty$~
\gr, as is seen from the asymptotic form
$$ K_\n (z) \approx {\sqrt{ \fracm \pi{2z} }} e^{-z} (1+{\cal O}(z^{-1}))
{}~~~~(z\approx \infty) ~~.
\eqno(3.15) $$

We next study the dilaton field equation (3.2) for the solution (3.12).
Now due to $~\varphi=0$,
$$ r^{-2a} (\partial_x A_1)^2 + r^{-2a} (\partial_y A_1)^2 + M^2 A_1^2
= r^{-2a} \a^2 \left(\partial_r K_0(\z) \right)^2 + M^2 \left(
K_0(\z)\right)^2 = 0 ~~
\eqno(3.16) $$
should vanish.  In fact, each term here vanishes desirably, when we take
the limit
$$ M\rightarrow \infty ~~,
\eqno(3.17) $$
because of the relations from (3.15) such as
$$\eqalign{r^{-a}\partial_r K_0(\z)
&\approx - M K_0(\z) \cr
&\approx - M {\sqrt{\fracm{\pi(a+1)}{2M r^{a+1}}}} \exp\left(-\fracm M{a+1}
r^{a+1} \right) \left[ \,1+{\cal O}(M^{-1}) \,\right] \rightarrow 0
{}~~~~(M\rightarrow\infty) ~~. \cr }
\eqno(3.18) $$

The $~G\-$field eq.~(3.3) vanishes, due to the vanishing
$~B\-$fields.  The remaining field equation is (3.1), which is
re-written in terms of the polar coordinates $~(r,\theta)$~ as
$$ \li{&R_{r r} =  2\left( \partial_r^2 \s + \fracmm 1 r \partial_r
\s\right) = 2 A_1''(r) + 4 M^2 r^{2a} (A_1(r))^2 ~~,
&(3.19) \cr
& R_{\theta\theta} = r^2 R_{r r} ~~.
&(3.20) \cr } $$
Recall (3.7) and the property of the Green's function
\ref\fms{{\it See e.g.}, D.~Friedan, E.~Martinec and S.~Shenker,
\np{271}{86}{93}.} that
$$ \partial_r^2(\ln r) + \fracmm 1 r
\partial_r(\ln r) = - \fracmm {\d(r)} r ~~.  \eqno(3.21) $$
Using these in the middle side of (3.19) implies that
$$ - 2a \fracmm {\d(r)} r = 2 A_1''(r) + 4M^2 r^{2a} (A_1(r))^2 ~~
\eqno(3.22) $$
is to be satisfied.  In fact, the r.h.s.~of (3.22) vanishes everywhere in
$~r>0$~ under $~M\rightarrow\infty$, and the singularity at $~r=0$~ can be
estimated by the integration  $$\eqalign{&\int_0^\infty d r \, r \, \left[ 2
A_1''(r) + 4M^2 r^{2a}  (A_1(r))^2
\right] \cr & ~~~ = 2\a^2 (a+1)^2 \int_0^\infty d\z\, \z \left[ (K_0'(\z))^2 +
2
(K_0(\z))^2   \right] = 2\a^2 (a+1)^2 (c_1 + 2c_0) ~~,  \cr }
\eqno(3.23) $$
where we have used the relation $~K_1'(\z) = - K_0(\z)$, and $~c\,'s$~ are
positive constants:
$$ c_0 \equiv \int_0^\infty d\z\, \z (K_0(\z))^2 ~~, ~~~~
c_1 \equiv \int_0^\infty d\z\, \z (K_1(\z))^2 ~~.
\eqno(3.24) $$
Comparing (3.23) with the l.h.s.~of (3.22) yields the condition
$$ \a = \pm\fracmm 1{|a+1|} {\sqrt{ \fracm{-a}{2c_0+c_1} }}~~~~~(-1<a<0)~~.
\eqno(3.25) $$
Note that $~a<0$~ is to be satisfied for this to make sense.  Eventually,
we have established that
$$ R_{r r} = - 2a \fracmm{\d(r)}r ~~,
\eqno(3.26) $$
which in turn implies the satisfaction of (3.20) due to $~r \d(r) \equiv 0$.

We finally inspect whether the Killing spinor equations $~\hbox{(2.4b)} =
\hbox{(2.4d)} = \hbox{(2.5c)} = 0$~ hold.  We start with (2.5c)
simplified as
$$  \left[ (\s_1 \e^i) \cos\theta + (\s_2\e^i) \sin\theta \right]
(r^{-a} \partial_r K_0(\z)) + M \e^i K_0(\z) {\eqques} 0  ~~,
\eqno(3.27) $$
for the representation $~\g^{(1)} = i \s_1,~\g^{(2)} = i \s_2,~
\g^{(0)} = -\s_3$.  We already know that each of these terms vanishes by
(3.18) when $~M\rightarrow\infty$.\footnotew{We can also make sure the
absence of the  $~\d$-function singularity at $~r=0$~ by an
$r$-integration.}  Eq.~$\hbox{(2.4d)} \eqques 0$~ is trivial due to
the  vanishing $~\varphi~$ and $~G$, while the only remaining one is
$~\hbox{(2.4b)} \eqques0$:
$$ D_m(e) \e^i \eqques 0 ~~,
\eqno(3.28) $$
which is sufficient for the vanishing commutator
$$ \[ D_{(1)}(e), D_{(2)}(e) \] \e^i =  - \fracm i 4 (\s_3 \e^i) (
g^{11} R_{11} + g^{22} R_{22} ) =  \fracm i2 a (\s_3 \e^i) r^{-2a-1} \d(r)
\eqques 0 ~~.
\eqno(3.29) $$
The Killing spinor equation (3.28) has a local solution,
when (3.29) vanishes under the condition ~$ - 2 a - 1 > 0$, {\it i.e.},
$~a < - 1/2$~ consistently with (3.13) and (3.25), so eventually
$$ -1 < a < - \half ~~.
\eqno(3.30) $$
Since no integer is allowed for $~a$, our mechanism automatically generates
the conical singularity \djt.  Note also that the case $~a=-1$~ would
correspond to {\it no} conical singularity.

The satisfaction of Killing spinor equations above, however, is
formal in the sense that there is actually no covariantly constant spinors
that satisfy the global boundary condition in a conical space-time \witten.
This can be most easily seen
by solving eq.~$\hbox{(3.28)}= 0$:  Using $~\omega\du1{(1)(2)} = - a
y/r^2, ~\omega\du2{(1)(2)} = a x/r^2$, we get
$$ \partial_x \e^i - \fracmm{i a y}{2r^2} \s_3\e^i= 0 ~~, ~~~~
\partial_y \e^i + \fracmm{i a x}{2r^2} \s_3\e^i = 0 ~~,
\eqno(3.31) $$
which has only the $~r\-$independent solution
$$\e^i (\theta) = \pmatrix {e^{- i a\theta/2} & 0 \cr
0 & e^{+i a\theta/2} \cr} \e^i(0) ~~.
\eqno(3.32) $$
However, this solution does {\it not} satisfy the
usual periodic boundary condition for a fermion: $~\e(4\pi) \neq \e(0)$, and
therefore the such a covariant spinor does not globally exist.  Relevantly,
the $~N=2$~ supersymmetry is broken down to $~N=0$, and the degeneracy
between fermions and bosons will be shifted \witten\wittenduality, because
any $~\theta\-$dependent fermionic solution violates this boundary condition.
In ref.~\bbs\ there was a minimal coupling of an $~U(1)$~ gauge field $~A_\m$~
in a vector multiplet coupled to the gravitino which compensated the phase
discrepancy.  In our system a minimal coupling of $~B_\m$~ induces a
dilaton-dependent {\it negative} definite potential which becomes zero at
$~|\varphi| = \infty$~ with otherwise anti-de Sitter space-time
destabilizing the system.\footnotew{This situation is usual for automorphism
group gauging of supergravities with the gauge field {\it inside} the
supergravity multiplet, such as the gauged  $~D=4,\,N=8$~ supergravity
\ref\nicolai{B.~De Wit and H.~Nicolai, \pl{108}{82}{285}.}, {\it etc.}}  On
the other hand, even though the minimal coupling of  a vector field
$~A_\m$~ (instead of $~B_\m$) in an independent vector multiplet may have
most likely a positive definite potential, we doubt its validity due to
inconsistency between the $~\psi\partial\varphi$~ and ~$\psi F\-$sectors
after the $~\Tilde g\-$dependent variations of candidate lagrangian
terms,\footnotew{Here $~\psi$~ is the
gravitino field and ~$\Tilde g~$ is the minimal coupling constant.} as
simple computations reveal.

To conclude, there is no surviving supersymmetry on this $~D=3,\,N=2$~
background, because the global periodic boundary condition for the
supersymmetry parameter is violated, even though the Killing spinor
equations are locally satisfied.  However, the cosmological constant still
vanishes, since the metric is asymptotically Minkowskian.

\bigskip\bigskip\bigskip

%\newpage

\centerline{\bf 4.~~Conical Singularity in $~D=4$}

Once we have understood the mechanism of generating a conical singularity
in $~D=3$, our next natural step is its
application to \dfno supergravity, which is of much more crucial
interest for phenomenological model building.

There are two options for the \dfno supergravity, either with or without
gauging the global axial $~U(1)_{\rm A}\-$symmetry.  In this paper we
consider the latter, because the $~U(1)_{\rm A}\-$guage field
will not play any important role for our purpose.

Our field content is the supergravity multiplet \dfno supergravity
$~(e\du\m m, \psi_\m)$~ coupled to $~n$~ copies of chiral multiplets
$~(\phi^i, \chi^i)~\,{\scst (i,~j,~\cdots~=~1,~2,~\cdots,~n)}$.
In this section we follow the notation of ref.~\ref\bagger{J.A.~Bagger,
\np{211}{83}{302}.} based  on the Euclidean signature $~(+,+,+,+)$~ in
ref.~\ref\pvn{P.~van Nieuwenhuizen,  \prep{68}{81}{189}.}, except for the
local Lorentz indices $~{\scst a,~b,~\cdots~=~(1),~(2),~(3),~(4)}$.
The total lagrangian we need is the sum of lagrangians (32) + (42) in
ref.~\bagger\ in the case of trivial target space dictated by eqs.~(47) -
(50) therein.  The purely bosonic part of the total lagrangian is
\bagger:
$$ \eqalign{e^{-1} \Lag_{\rm B} =  \, & - \half R - g^{\m\n}(\partial_\m
\phi^i) (\partial_\n\phi^{*i}) \cr
& + e^K \left[ \, 3|W|^2 - (D_i W) (D_{i^*} W^*) \,\right] ~~, \cr  }
\eqno(4.1) $$
where $~g$~ is the $~U(1)_{\rm A}$~ coupling constant, and
$$ D_i W \equiv W_{,i} + K_{,i} W \equiv \fracmm{\partial W}{\partial\phi^i}
+ \fracmm{\partial K}{\partial\phi^i} W ~~.
\eqno(4.2) $$
As in the $~D=3$~ case, we fix the K\"ahler and holomorphic potentials as
$$ K\equiv \sum_{i=1}^n \phi^{*i} \phi^i ~~, ~~~~  W = + \half\sum_{i=1}^n
m_i (\phi^i)^2 =  + \half M(\phi^1)^2 + \half\sum_{j=2}^n m_j (\phi^j)^2~~,
\eqno(4.3) $$
where as before we take $~M\rightarrow\infty$~ limit at the end.
It turns out that there is no need to add a constant to $~W~$ to fine-tune
the cosmological constant to be zero.\footnotew{In this section we do not
address ourselves to the problem with the fine-tuning.  (Cf.~section 5)}  In
fact, in terms of $~\phi^i \equiv (A_i + i B_i)/\sqrt2$~ the bosonic
potential is
$$ \li{&V(A_1,B_1, A_2,B_2, \cdots, A_n,B_n) \cr
& = \exp\left({\scst\sum_k} |\phi^k|^2\right)
\left[ \left| M\phi^1 + \phi^{*1} W \right|^2  + \sum_2^n \left|  m_j
\phi^j + \phi^{*j} W \right|^2 - 3|W|^2 \right]  ~~.
&(4.4) \cr} $$
Keeping only the terms up to the cubic, we can easily
show that this potential is minimized at its zero value for the v.e.v.'s
$~\phi^i = 0$.  If we set non-zero v.e.v.'s only for $~A_1$, then
$$ V(A_1,0,0,0, \cdots, 0,0) = \half M^2 A_1^2 e^{A_1^2/2}
\left[ 1 + \fracm 1 8 A_1^2 + \fracm1 {16} A_1^4 \right]
= \fracm 1 2 M^2 A_1^2 + {\cal O} (A_1^3) ~~.
\eqno(4.5) $$

We can now get all the relevant bosonic field equations
$$ \li{& R_{\m\n} = -
(\partial_\m A_1) (\partial_\n A_1) - (\partial_\m B_1)
(\partial_\n B_1) ~~,
& (4.6) \cr
& e^{-1} \partial_\m \left( eg^{\m\n} \partial_\n A_1 \right) -
\fracmm{\partial}{\partial A_1} V(A_1, 0, 0,0, \cdots, 0,0) = 0 ~~,
&(4.7) \cr
& e^{-1} \partial_\m \left( eg^{\m\n} \partial_\n B_1 \right) -
\fracmm{\partial}{\partial B_1} V(A_1, B_1, 0,0, \cdots, 0,0) = 0 ~~,
&(4.8) \cr } $$
where we are considering the case $~\phi^j =0~~{\scst
(j~=~2,~\cdots,~n)}$, because $~V$~ is minimized at these v.e.v.'s,
while keeping only $~A_1$~ to be non-trivial.

We now fix the ansatz for the vielbein for $~(x^\m) = (x,y,z,t)$~ as
$$ (e\du\m a) = \pmatrix{e^\s & 0 & 0 & 0 \cr
0 & e^\s & 0 & 0 \cr 0 & 0 & 1 & 0 \cr 0 & 0 & 0 & 1 \cr } ~~, ~~~~
\s = a \ln \r~~, ~~~~ \r \equiv {\sqrt{x^2 + y^2}} ~~,
\eqno(4.9) $$
like the $~D=3$~ case.  As for $~A_1$, we use exactly the same solution
as $~D=3$:
$$ A_1(\r) = \a K_0 (\zeta) = \a K_0 \left( \fracm M{a+1} \r^{a+1}
\right) ~~, ~~~~\zeta\equiv M (a+1)^{-1}\r^{a+1}~~.
\eqno(4.10) $$

We easily see that eq.~(4.8) is satisfied, because the first term vanishes
for $~B_1=0$, and we also know that the potential is minimized at $~B_1=0$.
Now eq.~(4.7) is re-written as
$$ \r^{-2a} \left(\partial_\r A_1 + \fracmm 1 \r
\partial_\r A_1 \right) - M^2 A_1 + {\cal O}(A_1^2) \eqques 0 ~~,
\eqno(4.11) $$
by (4.5).  This is satisfied as (3.16), because $~{\cal
O}(A_1^2)$~ damps much faster than other terms as $~M\rightarrow\infty$.

The remaining field equation to be examined is (4.6).
All the terms are of the same pattern as the $~D=3$~ case,
and the rest boils down to an analog of (3.25) {\it except for}
$~2c_0$~ in (3.25) replaced by $~c_0$~ now: $$ \li{\int_0^\infty d\r \, \r
R_{\r\r} = &\, 2 \int_0^\infty d\r \, \r  \left(\partial_\r ^2\s + \fracmm1
\r \partial_\r \s \right) \cr  = &\, \int_0^\infty d\r\, \left[ (\partial_\r
A_1)^2  + M^2 \r^{2a} A_1^2 \right] = \a^2 (a+1)^2 (c_0 + c_1) ~~,
&(4.12) \cr
\a = &\, \pm\fracm1{|a+1|} {\sqrt{ \fracm{-a}{c_0+c_1} }}~~~~~(-1<a<0)~~.
&(4.13) \cr } $$
This concludes the satisfaction of all the bosonic field equations
after taking the limit $~M\rightarrow\infty$, and the resemblance
between the $~D=4$~ and $~D=3$~ cases is clear.

We finally analyze the  Killing spinor equations by the
supertranslations of fermions \bagger:
$$\li{& \d\low Q \psi_\m = 2 {\cal D}_\m \e + \half \left( K_{,i}
\partial_\m \phi^i - K_{,i^*} \partial_\m \phi^{*i} \right) \g_5 \e
+ \half e^{K/2} W \g_\m \e ~~,
&(4.14) \cr
& \d\low Q \chi^i ={\sqrt 2} \g^\m \e \partial_\m\phi^{*i} - {\sqrt 2}
e^{K/2} \e D^i W ~~.
&(4.15) \cr } $$
These equations resemble the $~D=3$~ case.
Eq.~$\hbox{(4.15)}=0$~ has the linear terms similar to (3.27), and
higher order terms, both vanishing at $~M\rightarrow\infty$.

The non-trivial Killing equation is $~\hbox{(4.14)} = 0$, which has
a new term of $~W$~ compared with (3.31).  However,
this term is at least bilinear in $~A_1$~ which is easily shown
to vanish under $~M\rightarrow\infty$.  In fact, $~\hbox{(4.14)} = 0$~ is
solved by
$$\e(\theta) = I_2\otimes\pmatrix{e^{-ia\theta/2} & 0 \cr
0 & e^{+ia\theta/2} \cr } \e(0) ~~,
\eqno(4.16) $$
for $~\theta\equiv \arctan(y/x)$.  Corresponding to (3.29)we have the
condition
$$ \eqalign{\[ D_{(1)}(e), D_{(2)}(e) \] \e &\, =  - \fracm i 4
(I_2\otimes\s_3) \e ( g^{11} R_{11} + g^{22} R_{22} )  \cr
&\, = \fracm i2 a (I_2\otimes\s_3) \e \r^{-2a-1} \d(\r) \eqques 0
{}~~, \cr}
\eqno(4.17) $$
yielding
$$ -1 < a < -\half ~~.
\eqno(4.18) $$
The representation we have used is $~\g^{(4)} = -
\s_1 \otimes I_2 ,~\g^{(i)} = + \s_2 \otimes \s_i ~~{\scst(i~=~1,~2,~3)},
{}~\g_5 = + \s_3 \otimes I_2$.  Unlike that of $~D=3$~ in ref.~\bbs, we have
no $~U(1)_{\rm A}$~ coupling to $~\e$, so that we have no cancellation
between the Lorentz connection and $~A_\m$.\footnotew{Even if we introduce an
$~U(1)_{\rm A}$ minimal couplings, there will be no cancellation due to the
difference in the $~\g$-matrix structure in $~D=4$.}
Therefore as our previous
$~D=3$~ case, the covariantly constant spinor can exist only locally
but not globally, violating the boundary condition for a fermion:
$~\e(4\pi)\neq\e(0)$.  Accordingly, the original degeneracy between fermions
and bosons under supersymmetry such as the masses or couplings will be
lifted \witten\wittenduality, which may be useful for phenomenological
applications \ref\nilles{{\it See e.g.}, H.P.~Nilles, \prep{110}{84}{1}.}.
However, the vanishing of the cosmological constant is not disturbed by this
supersymmetry breaking, owing to the local satisfaction of bosonic field
equations, and the metric is asymptotically Euclidean.

\bigskip\bigskip\bigskip

\centerline{\bf 5.~~Duality between $~D=4$~ and $~D=3$}

We mention briefly the strong coupling duality \wittenduality\ between the
two systems in $~D=4$~ and $~D=3$, that may explain the automatic
zero-ness of the cosmological constant with broken supersymmetry in the
former.

As eq.~(2.7) for $~a=b$~ indicates, the field redefinitions
$$ \eqalign{& (\Tilde e\du\m m, \Tilde \psi_\m, \Tilde B_\m, \Tilde\chi)
\equiv (e^{-b\varphi} e\du\m m, e^{-b\varphi/2} \psi_\m,
B_\m, e^{+b\varphi/2} \chi) ~~, \cr
&(\Tilde A_a, \Tilde B_a, \Tilde\l_a ) \equiv (e^{-b\varphi/2} A_a,
e^{-b\varphi/2} B_a, e^{+b\varphi/2}\l_a) ~~, \cr }
\eqno(5.1) $$
make the dilaton-dependence manifest as the string coupling constant
\wittenduality:
$$\Lag_{\rm SG} = e^{b\varphi} \Tilde\Lag_{\rm SG} ~~, ~~~~
\Lag_{\rm SM} = e^{b\varphi} \Tilde\Lag_{\rm SM}  ~~, ~~~~
\Lag_m = e^{b\varphi} \Tilde\Lag_m\Big|_{m_a\rightarrow
e^{-b\varphi} \Tilde m_a} ~~,
\eqno(5.2) $$
where the {\it tilded} lagrangians have only the {\it tilded} fields.  In
$~\Lag_{\rm SG} + \Lag_{\rm SM}$, the dilaton without derivative appears
only in the over-all factor $~e^{b\varphi}$.
In $~\Lag_m$, the old masses $~m_a$~ are also formally replaced by
$~e^{-b\varphi}\Tilde m_a$.  From
$~\Lag_{\rm SG} + \Lag_{\rm SM}$~ the string coupling constant $~\l$~ is
identified as $~\l^{-2} = e^{b\varphi}$~ \wittenduality.

We can relate this result to the \dfno supergravity plus chiral multiplet
by the dimensional reduction on $~\IR^3\otimes S^1$~ with a
circle $~S^1$~ of radius $~r=e^{-b\varphi}$~ as\footnotew{Here we switch
to the signature $~(+,-,-,-)$.}
$$ d\Hat s{\,}^2 = \Hat g _{\hat\m\hat\n} d{\Hat x}^{\hat\m} d{\Hat
x}^{\hat\n} = e^{2b\varphi} \Tilde g_{\m\n} d x^\m d x^\n -
e^{-2b\varphi} (d {\Hat x} {}^3)^2 ~~,
\eqno(5.3) $$
where all the {\it hatted} quantities are for $~D=4$, while the {\it tilded}
$~{\Tilde g}_{\m\n}$~ accords with (5.1).  Accordingly, the \dfno
supergravity multiplet $~(\Hat e\du{\hat\m}{\hat m},\Hat\psi_{\hat\m})$~ is
reduced to the \dtnt supergravity $~(e\du\m m,\psi\du\m
i,B_\m,\chi^i,\varphi)$, while the chiral multiplet $~(\Hat\phi_i,
\Hat\chi_i)$~ is reduced to the scalar multiplet
$~(A_a,B_a,\l_a^i)$.\footnotew{The index-conventions follow
the corresponding section of each multiplet.}
Now from the \dfno lagrangian \bagger\ of section 4, we can identify the
radius \wittenduality   $$ r = e^{-b\varphi} = \l^2 ~~,
\eqno(5.4) $$
and its bosonic part\footnotew{Due to supersymmetry,
the same is also true for the fermionic part.} after this dimensional
reduction agrees with that of $~\Lag_{\rm SG}+\Lag_{\rm SM}$~ in (5.2)
constructed within $~D=3$.

This relationship satisfies the criterion of consistency for duality in
ref.~\wittenduality, namely if the strong coupling limit $~\l\rightarrow
\infty$~ in $~D=3$, then the $~S^1$~ radius $~r\rightarrow\infty$,
implying the promotion (or ``oxidation'') of the $~D=3$~ system to a $~D=4$~
system with the Poincar\'e invariance.  Since $~e^{b\varphi} m_a =
m_a/\l^2$, our limiting procedure $~m_1\equiv M\rightarrow\infty$~ may well
be interpreted as the effective mass $~\Tilde M\equiv M/\l^2$~ kept finite
in $~D=4$.  Now the question of the fine-tuning of the $~D=4$~ cosmological
constant is solved, due to the fundamental $~D=3$~ theory automatically
yielding the zero cosmological constant under this duality,
absorbing the undesirable massless dilaton at the same time
\wittenduality.

\bigskip\bigskip\bigskip

\centerline{\bf 6.~~Concluding Remarks}

In this paper we have presented a mechanism of generating a conical
singularity in $~D=3$~ directly applicable to $~D=4$.
As a by-product, \dtnt supergravity lagrangians are presented in terms of
the supergravity multiplet with a dilaton and an antisymmetric tensor
coupled to massive scalar multiplets, which have not been presented
elsewhere to our knowledge.

We have also seen that the strong coupling duality relation outlined in
\wittenduality\ is realized between the systems of two dimensionalities,
namely when the string coupling constant $~\l$~ in $~D=3$~ is taken to
infinity, the $~S^1$~ radius for the compactification from $~D=4$~ grows.
This implies that the $~\l\rightarrow\infty$~ limit in $~D=3$~ is
equivalent to a $~D=4$~ system with the Minkowskian metric and Poincar\'e
invariance.  To put it differently, this duality can be regarded as
dimensional ``oxidation'' from $~D=3$~ to $~D=4$~ which is the reversed
process of the usual dimensional ``reduction'' from $~D=4$~ to $~D=3$.
Accordingly, the supersymmetry is broken in $~D=4$~ with a conical
singularity, but the zero-ness of the cosmological constant is automatically
realized without fine-tuning by this duality.  Our peculiar limit
$~M\rightarrow\infty$~ we adopted in our mechanism maintains the finite
effective mass in the resulting $~D=4$.

The limit $~M\rightarrow\infty$~ is also natural from another viewpoint.
For example, the point-mass source in ref.~\djt\ creating conical
singularities should correspond to the point particle limit in a field
theory, and such a limit must be realized by taking the mass of a particle
to infinity in order to enhance the ``point particle'' effect,
suppressing the ``wave'' effect.  The $~M\rightarrow\infty$~ may well be
related to the familiar limit $~M_{\rm Pl}\equiv1/\k\rightarrow
\infty$~ \nilles\ for the ``low energy'' physics such as local supersymmetric
grand unifications below the Planck mass $~M_{\rm Pl}$.  Note also that
important relations in our procedure such as the condition (3.30) or (4.18)
do not depend on the value of $~M$, so that the $~M\rightarrow\infty$~ limit
makes sense.  This non-trivial feature indicates that our
$~M\rightarrow\infty$~ limit has some fundamental significance controlling
the conical singularity of the space-time in a ``topological'' way.  It is
interesting that our mechanism has a finite breaking effect
by $~a\approx{\cal O}(1)$~ instead of $~M/M_{\rm Pl} \rightarrow 0$~
\bbs\ well below $~M_{\rm Pl}$.

\bigskip\bigskip

We are grateful to S.J.~Gates, Jr.~for helpful suggestions and
encouragement.

\bigskip\bigskip\bigskip

\vfill\eject

\footatend\vfill\supereject\immediate\closeout\rfile\writestoppt
\baselineskip=14pt\centerline{{\bf References}}\bigskip{\frenchspacing%
\parindent=20pt\escapechar=` \input refs.tmp\vfill\eject}\nonfrenchspacing

\end{document}